\documentclass{emulateapj}
\usepackage{apjfonts}
\usepackage{amsmath}
\usepackage{psfig}
\usepackage{nicefrac}
\def\gs{\mathrel{\raise0.35ex\hbox{$\scriptstyle >$}\kern-0.6em 
\lower0.40ex\hbox{{$\scriptstyle \sim$}}}}
\def\ls{\mathrel{\raise0.35ex\hbox{$\scriptstyle <$}\kern-0.6em 
\lower0.40ex\hbox{{$\scriptstyle \sim$}}}}

\def\etal{\hbox{et al.}}
\def\OII{\hbox{[O II]}$\,\,$}
\def\Hd{\hbox{H$\delta$}$\,\,$}
\def\Hb{\hbox{H$\beta$}\,\,}
\def\Ha{\hbox{H$\alpha$}$\,$}

\def\Msun{\rm{\hbox{M$_{\odot}$}}}           
\def\ang{\hbox{$\,$\AA}}
\def\kms{\rm{\hbox{km s$^{-1}$}}}
\def\OII{\hbox{[O II]}$\,$}
\def\OIII{\hbox{[O III]}$\,$}

\def\Hd{\hbox{H$\delta$}$\,$}
\def\24m{\hbox{24~$\micron$}$\,$}
\def\10-18{\hbox{$\times~10^{-18}$}}
\def\deg{\hbox{$^{\circ}$}}

\def\lya{\mbox {Ly-$\alpha$~}}
\def\Lya{\mbox {Ly$\alpha$}}
\def\flux{~ergs~s$^{-1}$~cm$^{-2}$}
\def\lum{~ergs~s$^{-1}$}
\def\kms{~km~s$^{-1}$~}

\slugcomment{Submitted: 2014 September 15; accepted:  2015 April 11}

\shortauthors{Dressler \etal\ }
\shorttitle{Faint \Lya\ Emitters Confirmation}

\begin{document}

\title{Confirmation of a Steep Luminosity Function for \Lya\ Emitters At \lowercase{z} = 5.7: A Major Component of Reionization\footnote{T\lowercase{his 
paper includes data gathered with the 6.5 meter \uppercase{M}agellan \uppercase{T}elescopes located at \uppercase{L}as 
\uppercase{C}ampanas \uppercase{O}bservatory, \uppercase{C}hile.}}}

\author{Alan Dressler}
\affil{Carnegie Observatories, 813 Santa Barbara St., Pasadena, California 91101-1292}
\email{dressler@obs.carnegiescience.edu}

\author{Alaina Henry}
\affil{Astrophysics Science Division, Goddard Space Flight Center, Code 665, 
Greenbelt, MD 20771}
\email{alaina.henry@nasa.gov}

\author{Crystal L. Martin}
\affil{University of California, Santa Barbara, Department of Physics, Santa Barbara, CA 93106}
\email{cmartin@physics.ucsb.edu}

\author{Marcin Sawicki}
\affil{St. Mary's University, Department of Astronomy and Physics, 923 Robie Street,
Halifax, N.S., B3H 3C3, Canada}
\email{sawicki@ap.smu.ca}

\author{Patrick McCarthy}
\affil{Carnegie Observatories, 813 Santa Barbara Street, Pasadena, California 91101-1292}
\email{pmc2@obs.carnegiescience.edu}

\author{Edward Villaneuva}
\affil{Carnegie Observatories, 813 Santa Barbara Street, Pasadena, California 91101-1292}
\email{edwardv@obs.carnegiescience.edu}

\begin{abstract}

We report the first direct and robust measurement of the faint-end slope of the Ly-$\alpha$ emitter (LAE) 
luminosity function at z = 5.7. Candidate LAEs from a low-spectral-resolution blind search with IMACS on 
Magellan-Baade were targeted at higher resolution to distinguish high redshift LAEs from foreground galaxies.  
All but 2 of our 42 single-emission-line systems have flux $F < 2.0 \times 10^{-17}$\flux, making these 
the faintest emission-lines observed for a $z = 5.7$ sample with known \emph{completeness}, an essential 
property for determining the faint end slope of the LAE luminosity function.  We find 13 LAEs as compared to 
29 foreground galaxies, in very good agreement with the modeled foreground counts predicted in Dressler 
\etal\ (2011a) that had been used to estimate a faint-end slope of $\alpha = -2.0$ for the LAE luminosity 
function.  A 32\% LAE fraction, LAE/(LAE+foreground), within the flux interval $F = 2-20\,\, \10-18$ \flux, 
constrains the faint end slope of the luminosity function to $-2.35 < \alpha < -1.95$ (1$\sigma$).  We show 
how this steep LF should provide, to the limit of our observations, $M_{UV} \sim$ -16, more than 20\% of the 
flux necessary to maintain ionization at $z=5.7$, with a factor-of-ten extrapolation in flux reaching more than 50\%.  
This is in addition to a comparable contribution by brighter Lyman Break Galaxies $M_{UV} \ls$ -18. We 
suggest that this bodes well for a sufficient supply of Lyman continuum photons by similar, low-mass star 
forming galaxies within the reionization epoch at $z \approx7$, only 250 Myr earlier.
 

\end{abstract}


\keywords{galaxies: high-redshift -- galaxies: evolution -- galaxies: formation}

\section{Introduction}

Our understanding of galaxy evolution during the epoch of reionization has improved with the deep near-IR
imaging from WFC3 on the Hubble Space Telescope.  Numerous \emph{Lyman-break} galaxies (hereafter, LBGs) 
have been found at redshifts $z=6-9$,  with a \emph{luminosity function} (hereafter, LF) that spans a factor of 
$\sim$100 in brightness (e.g., McLure \etal\ 2013;  Ellis \etal\ 2013; Bouwens \etal\ 2014; Oesch \etal\ 2014).   
Although the photometric redshifts of these young galaxies are reasonably secure, spectroscopic confirmation of 
\Lya\ emission has proven elusive in most cases (Fontana \etal\ 2010; Pentericci \etal\ 2011; Schenker \etal\ 2012;
Caruana \etal\ 2012, 2014; Bunker \etal\ 2013).  There is mounting evidence that this is due to a significant fraction of 
remaining HI that substantially attenuated any \Lya\ emission escaping these young objects (Stark \etal\ 2010; 
Ono \etal\ 2012; Treu \etal\ 2013; Tilvi \etal\ 2013; Momose \etal\ 2014; cf. Dijkstra et al. 2014) .

Young stellar populations in early galaxies were the likely sources of high-energy (E > 13.6 eV) photons responsible 
for reionization of the intergalactic medium (IGM).  However, it is well known that the brighter LBGs, $L\gs L^*$, 
provide a small fraction of the required flux, so that much larger numbers of fainter, unobserved galaxies would be 
needed to balance or exceed  the ionizing budget (Bunker \etal\ 2010).    In fact,  recent surveys that reach deeper do 
suggest that the LBG LF \emph{is} steep, with faint-end slope $\alpha\sim$\, -2.0  (Bradley \etal\ 2012; Alavi \etal\ 
2014; Schmidt \etal\ 2014; Bouwens \etal\ 2014).  Though the \emph{observed} LBGs account for $\sim $10-20\% of the 
required Lyman-continuum (LyC) flux, if a slope of $\alpha\sim$\,-2.0 continues to a luminosity $M_{UV} \sim$ -13,  
then LBGs could account for all of the flux required for full reionization (Robertson and Ellis 2012;  Robertson 
\etal\ 2013; Schmidt \etal\ 2014; Robertson \etal\ 2015).  

Even with sufficient numbers, however, it is not certain that LBGs can supply sufficient LyC photons into the 
IGM: at redshifts $z=5-6$, where neutral hydrogen is gone from the IGM, \Lya\ emission is only sometimes
detected in LBGs (e.g., Shapley \etal\ 2003; Kornei \etal\ 2010; Stark \etal\ 2010; cf.~Curtis-Lake \etal\ 
2014).  Therefore, it is important to investigate the contribution of LyC photons by the   class of galaxies 
\emph{defined} by strong \Lya\ --- the \emph{\Lya-emitters} (LAE).  As described by Schaerer (2014), LAEs 
and LBGs at high redshift are closely related star forming systems whose differences in  observable properties 
could be due entirely to differences in dust content.  The lower (on-average) stellar mass of  LAEs compared to 
LBGs may be connected to their systematically lower dust contents.  It is possible, then,  that the mature stellar 
populations in LBGs, evident in their strong stellar UV-continua, entrain enough dust to prevent many \Lya\ 
photons, and most LyC photons, from leaving the galaxy.  For example, from observations comparing LBGs and 
LAEs at $z \sim 3$, Nestor \etal\ (2013) infer LyC escape fractions 2-4 times higher for LAEs.  For this  reason, the 
needed LyC photons may preferentially come from LAEs, where stellar continuum radiation is weak, and the 
dominance of emission is the signature of a younger starburst --- perhaps the first major episode of star formation in 
the system. 

The largest collections of LAEs at $z>5$ come from narrow-band imaging surveys with the Subaru telescope 
(e.g., Shimasaku \etal\ 2006; Ouchi \etal\ 2008; Hu \etal\ 2010; Kashikawa \etal\ 2011).  With the wide 
field-of-view of the SuprimeCam, narrow-band searches are an efficient way to find high redshift LAEs with 
luminosities $L\gs L^*$, $\approx10^{43}$\lum. Thus, these studies have a good purchase on two of the Schechter function 
parameters, $\Phi^*$ and $L^*$, within their significant covariance.  However, as explained in Dressler \etal\ (2011a), MNS2) 
and reiterated in \S3.2 of this paper, detections of LAEs below L* become rapidly incomplete for narrow-band observations of 
$\sim$150\ang\ FWHM.  Although some fainter objects are detected, incomplete sampling makes such data at $z=5.7$ 
unsuitable for measuring the faint-end slope $\alpha$.

Our \emph{Multislit Narrowband Survey}, hereafter MNS, was specifically designed to produce complete samples of
LAEs up to ten-times fainter than the narrow-band imaging surveys.  Basically, this is accomplished by searching the 
same low-OH-background part of the spectrum as for imaging surveys for $z=5.7$ LAEs ($8110-8270$\ang), but by adding
a grism or grating to disperse the light so that each emission-line-detection competes against a ten-times-lower sky background.  
The origin of the technique, and our application of it using the 27 arcmin-diameter field of the IMACS f/2 channel, is detailed in 
Martin \etal\ (2008, MNS1) and in MNS2.

MNS2 describes the analysis of an excellent observing run in 2008 that produced $\sim$20 hours of integration for each of 
two fields, netting a sample of 210 single-emission-line-only sources that were candidate LAEs at 
$z=5.7$.  These spectra reached a 50\% completeness at a line flux of $F = 3.5\10-18$ \flux, sufficiently faint for the first 
credible measurement of the faint-end-slope of the LAE luminosity function (hereafter, LF).  However, because of the relatively 
low spectral resolution of $\approx$10\ang\ FWHM and the $\sim$150\ang\ coverage  of the search spectra, LAEs could not 
be reliably separated from foreground galaxies producing [O II], [O III], H$\beta$, or H$\alpha$ emission, which together were 
expected to outnumber the LAEs by about 2-to-1.  In MNS2 we used published results of counts of these foreground sources 
--- extrapolated to the fainter limits of the MNS survey --- to statistically correct for the foreground contamination and construct 
the residual LAE LF.  This process depended most sensitively on the faint-end slope of each of the foreground populations, 
whose value and range we needed to estimate.  Our best estimates of these quantities led to a faint-end slope of the LAE LF 
of $\alpha \approx$ -2.0, but values as low as -1.5 or as high as -2.5 could not be ruled out.

Confirmation of a steep slope for LAEs at $z\sim6$ has important implications for questions of galaxy formation, the
production of heavy elements in the universe, and reionization, so we have been strongly motivated to confirm the result
of a steep slope of the LAE LF forecast by our statistical correction for foreground contamination.  Accomplishing this
requires higher dispersion spectra for a statistically significant sample of the faintest LAE candidates.  Our 
first efforts to do this have been described in Henry \etal\ (2012, MNS3), where LAEs were positively identified in the 
COSMOS field using spectra from Keck-DEIMOS with a resolution of $\lambda\approx2$\ang; these results are briefly
reviewed in \S4.1.  In this paper we present similar spectra for a significantly larger sample of faint LAEs in our 15h field 
(LCRIS), leading to a determination $\alpha=-2.15\pm0.20$, in good agreement with the results of MNS2.

The paper is organized as follows: \S2 describes the new data taken with IMACS on Magellan;  \S3 describes how these 
objects were matched to those found in the low-resolution search, and the criteria for separating LAEs from 
foreground galaxies;  \S4 explains how we used these data to constrain the faint-end slope of the LAE LF; \S5 explores
the implication of this now-confirmed steep slope of the LAE LF for reionization; and \S6 gives our conclusions.

We adopt cosmological parameters of $\Omega_m = 0.30$, $\Omega_{\Lambda} = 0.7$, and $H_{0}  = 70$\kms 
Mpc$^{-1}$ throughout.

\section{The Data: Higher Dispersion Spectra of candidate LAEs}

The experimental technique of the 2008 MNS  search was to use 100 parallel  long slits crossing the full field of view 
of the IMACS f/2 camera, a circle of 27 arcmin diameter.  The spacing was chosen to allocate about $\sim$70 
pixels in the dispersion direction per slit, which covered a ``low-OH-emission" spectral band of $\lambda$ = 
8115 -- 8250\ang\ at 2.0\ang\ pix$^{-1}$.  The 2008 MNS  search used this setup, described in MNS2, to cover 
$\sim$55 sq arcmin ($\sim$10\% of the full f/2 imaging field) in both the COSMOS field and the Las Campanas 
Redshift Survey field (Marzke \etal\ 1999).  Slits 1.5 arcsec wide produced a spectral resolution of 14\ang\ for 
objects that fill the slits, but for the typical size and profile of the discovered single-line sources, and the good 
seeing conditions of the search ($<$0.6 arcsec FWHM) a 10\ang\ FWHM resolution was typical.  Still, at this 
resolution, spectra of LAEs are usually indistinguishable from single-emission-line  foreground galaxies, since 
a resolution of less than $\ls$5\ang\ is required to resolve the characteristic asymmetry of most \Lya\ emission 
lines, or to split the doublet of  \OII\ foreground sources at $z\approx1.2$.  Because these data could not be used
to unambiguously identify the LAEs at $z\approx5.7$, the result from MNS2 of a steep slope of $\alpha\approx-2.0$ 
for the LAE LF depended on a statistical correction for the foreground contamination.  Since this further depended 
on an extrapolation of the LFs for foreground sources to fainter limits than observed, the putative steep slope of MNS2
required further spectroscopy, to identify LAEs on an individual basis.

Such follow up spectral observations at 2-3\ang\ FWHM resolution were planned with both IMACS and 
Keck-DEIMOS starting in 2010.  Observations planned in 2010, 2011, and 2012 for IMACS on Magellan-Baade  
were thwarted by poor weather, but observations in 2010 and 2011 with DEIMOS of LAE candidates in the 
COSMOS field were moderately successful in terms of observing conditions.  The DEIMOS observations confirmed 
6 LAEs from the faint sample; the basic results of MNS3 are reviewed in \S4.1.  

In April 2013 and March 2014, two mostly-clear 5-night runs at Las Campanas Observatory, with average, 
on-target seeing of 0.68 arcsec and 0.71 arcsec (approximately the median seeing at Magellan), were successfully 
completed using IMACS in f/4 mode (Dressler \etal\ 2011b) with a 600-l/mm +13\deg\ blaze grating, delivering 
a scale of 0.378\ang\ pix$^{-1}$ and a spectral resolution (1.0 arcsec-wide slit) of $\sim$3\ang.  The detector 
readout was rebinned by a factor-of-two in the spatial direction to increase signal over read noise, resulting in a 
scale of 0.22 arcsec pix$^{-1}$.  A single slit mask was designed and fabricated for each year; each mask targeted 
LAE candidates in the 15h field of the \emph{Las Campanas Redshift Survey} (Marzke \etal\ 1999).  The position angle 
of slits was rotated by 90\deg\ from that of the long slits in the 2008 search mask, in order to place \emph{along} the slit 
the coordinate that includes a degeneracy between sky position and line wavelength. The multislit masks of the IMACS f/4 
cover a field of 15 arcmin x 15 arcmin. For the 2013 and 2014 runs, total integration times were 27.4 and 17.5 hours,  
respectively.  The spectral range extended out to 9000\ang\ for all spectra, and for most extended down to $\sim$6000\ang, 
important for confirming those cases where \Ha\ was the line detected in the search window.

The new IMACS f/4 spectra were reduced using the \emph{COSMOS} software package  ---
http://code.obs.carnegiescience.edu/cosmos/Cookbook.html, augmented by programs written in Python by
Kelson that facilitated the reduction of emission-line only sources, a departure from the common data 
reduction with  \emph{COSMOS} that makes use of object continua for fine-tuning object detection.
Wavelength calibration and registration were performed using He+Ne+Ar lamp spectra taken in proximity to 
each set of science frames, while the modeling and subtraction of sky was done using the Kelson (2003)
procedure.  The reductions produced 2D frames of sky subtracted spectra that were shifted and added 
using \emph{IRAF} `imcombine' to produce a single frame for each year's observations.  These were 
examined with \emph{Viewspectra}, a \emph{COSMOS} routine for interactive examination of 2D spectra 
and for extracting 1D spectra.


\begin{figure}[t]

\centerline{
\includegraphics[width=3.2in]{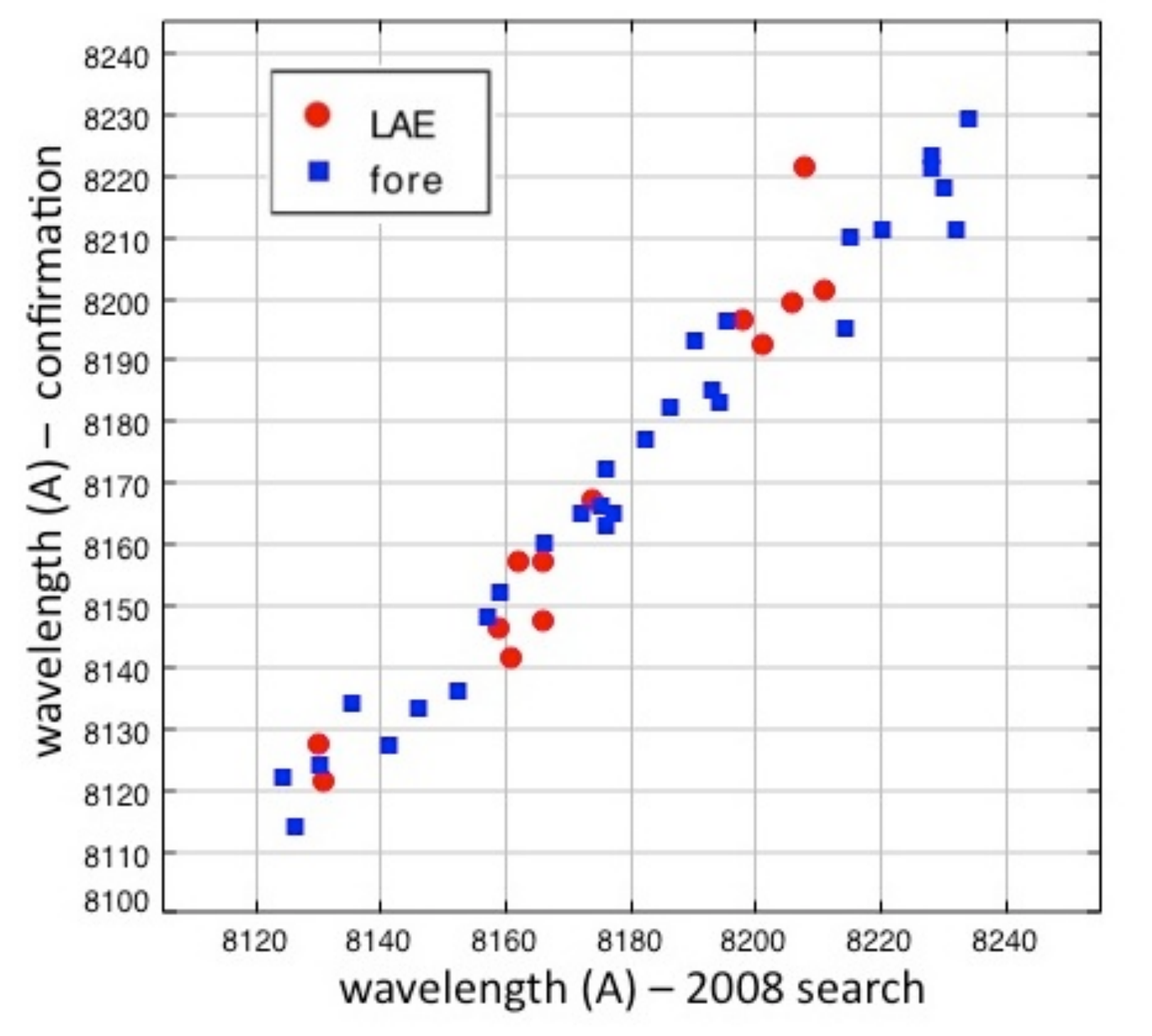}
}

\caption{Wavelengths of single-emission-line sources from the 2008 search for faint LAEs compared to 
confirmation observations with higher spectral dispersion in 2013 and 2014.  The dispersion of 5\ang\ and 
offset of 9\ang\ is dominated by uncertainties in the 2008 search data, for which wavelength calibration is 
difficult.  The apparent clumping of the LAEs into three systems with a typical separation of $\sim$200\kms is 
likely to be real and indicative of a significant cosmic variance.}


\end{figure}

Redshifts were measured for 45 of the 52 LAE candidates ("single-emission-line, no-blue-continuum sources" 
--- see MNS2), an 87\% success rate of recovering the targets from the 2008 search catalog.  The other 
13\% failed to show an emission-line at or near the predicted spatial or wavelength position; in fact, none 
of these showed a convincing line over the full wavelength band.   Twelve of the 52 candidates were repeated 
in the 2014 spectroscopy.  Nine of the repeats were recovered spectra in good agreement with the 2013 data, 
while for three no object was found, as in 2013.

\section{Results}

The final sample consists of 45 spectra: 13 sources are identified as LAEs and 32 as foreground galaxies
(identified through criteria described below). For 3 of  foreground galaxies the recovered emission line was an 
order-of-magnitude  brighter than than that of the 2008 candidate.\footnote{One of these was at the wrong 
spatial position  and another had a strong  continuum as well.}  These were judged to be cases where the LAE 
candidate was actually an H\,II region of a foreground galaxy, and the galaxy to which it belonged revealed when 
the slit orientation was changed by 90\deg\ to remove the wavelength-position ambiguity (see \S2).  Also, two of 
the foreground galaxies are excluded from the following analysis because, though they \emph{are} confirmations 
of the 2008 candidate data, they have fluxes of $F = 27 \,\,\&\  84\,\times 10^{-17}$\flux: this is a factor-of-ten brighter 
than the flux interval we are interested in.  The probability of finding a LAE this bright is less than 1\% for the area 
covered by our survey.  This leaves a sample of 13 LAEs and 27 foreground galaxies covering the range in flux 
$F$ = 2-20\,\10-18\flux, the relevant range for the determination of the faint-end  slope of the luminosity function 
(fainter than L$^*$), as can be seen in Figures 10 and 11 of MNS2. 

Concerning the measured fluxes of these sources, we note that photometry is problematical with spectroscopic 
data, due to the uncertainty of object position with respect to slit and the fact that some objects are bigger than 
the slit width (slit losses).  For the faint objects of our study, sky subtraction and flat-fielding errors add to the 
difficulty.  Furthermore, the detections at the flux limit of our sample, $F \sim 3\,\10-18$\flux , have only 
signal-to-noise ratio (hereinafter, SNR) $\approx3-5$ (see MNS2, Figure 3).  The combined  effect, evident in comparing 
our 2008, 2013, and 2014 data sets, is that photometry accurate to $\ls$10\% is not possible --- typical errors are typically 
two or three three  times larger.  In MNS3 we implemented a maximum-likelihood methodology capable of deriving the 
luminosity function in the presence of such photometric scatter and the uncertain positions of objects within the slits 
of the blind search.  In this paper we take an alternative approach based on the LAE fraction which, as we show in \S4, 
is also robust to these effects.  

We retained the  fluxes measured from the 2008 search spectra for the following analysis, however, among the 
27 foreground sources of the final sample (but for none of the LAE sample) 4 objects were significantly brighter 
in the 2013, 2014 spectroscopy than in the discovery spectra, 2 by $\sim$30\% and 2 by a full factor-of-two.  
We judged these to be cases of slit-losses in the discovery spectra --- a reasonable fraction --- and revised 
them to the higher value.

\subsection{Criteria for discriminating $z=5.7$ LAEs from foreground galaxies}

We have a high degree of confidence in the sample of LAEs we report here.  The criteria that  underlie this 
confidence is a series of qualifications.  To be considered a "recovered candidate" from the the 2008 
search, an emission line in the 2013 and/or 2014 spectra must agree with the wavelength found in the search
data.  Figure 1 shows this comparison for both the LAEs and foreground sources in the new data.  Compared
to the $\sim$135\ang\ range of the bandpass, the $\sim$5\ang\ scatter in the relation is small, ergo, there is
no question that the recovered objects are the ones found in the 2008 search.\footnote{The dispersion in
wavelength, as well as the $\sim$10\ang\ shift between the search data and the follow-up data is dominated
by the former.  Repeat measurements in 2014 of 12 objects observed in 2013 show a typical error of less than 
1\ang, from well-calibrated arc lines spanning of several thousand angstroms.  The are no comparison arc lines 
in the narrow band of the 2008 search data, which covers only $\ls$150\ang.  We used the narrow-band 
interval itself to define the wavelength scale, but the bandpass shifts with angle from the optical axis, and the 
"venetian blind" mask made used in the LCRIS (15h) search added additional uncertainty because of departure from 
sphericity of the highly perforated mask, another source of error in the wavelength.}  The recovered spectra were 
also required to lie within $\pm2$ arcsec  of the spatial position on the slit predicted from the 2008 search data.

Further criteria for identifying \Lya\ emission come from the spectra, most of which are shown in Figures 2 
and 5.   In Figure 2, the extracted spectra have been smoothed by gaussian of width $\sigma = 1.0$\ang\ 
(compared to the instrumental resolution of $\sim$2.5\ang) and plotted centered on the line detection, 
over an interval of 60\ang.   In Figure 5. in the Appendix, we show the left and middle panels of the Figure 
2 spectra over the full 135\ang\ bandpass and essentially unsmoothed, to allow a better judgement of the 
prominence of the detected lines and  the noise background.  Figure 5 also records the SNR for each of the 
detected lines, demonstrating  that the features are all detected at $SNR > 6 \sigma$.

The second criterion comes from the clear identification of foreground objects from their spectra.  \OII\ emission 
at $z\approx1.20$ accounts for $\sim$60\% foreground contamination.   As shown in the middle panel of Figures 2 
(and even more clearly in Figure 5), the \OII\ doublet ($\lambda\lambda$3726, 3729\ang) is well-resolved and each 
line easily distinguished,  even for the faintest objects.  Emission-line galaxies at $z\approx0.64$ are also a major 
component of the foreground: the $\lambda$5007 line of \OIII\ is shown for 5 out of the 8 cases  (right-hand column 
of Figure 2) and in all these cases $\lambda$4959 is also detected, and usually H$\beta$ as well.  \Ha\ at $z=0.25$ and 
\Hb\ at $z\approx0.68$ accounts for only 10\% of the foreground, and only one of these show accompanying [N II] 
emission, but in all but one case \Ha\ is ruled in or out by the detection of \OIII\ at an observer-frame wavelength of 
$\sim6250$\ang.  A comparatively rare [Ne III] line, confirmed by the presence of  \OII, was also found, but together 
\OII, \OIII/H$\beta$, and \Ha\ should account for 99\% of the foreground, since  these are much stronger than any 
other lines from \OII\ to \Ha.  H$\gamma$ or \Hd\ emission could  have been found, but \OII\ would always accompany 
them.   

The remaining 13 emission lines, shown in the left column of Figure 2, are identified as \Lya.  Although their principal
criterion is through elimination of other possibilities, there is additional verification from the typically 10-20\ang\
width of \Lya\ emission.  The  characteristic asymmetry of \Lya\ is seen in 8 of the 13 objects, and 3 others, although not 
clearly asymmetric at this SNR are clearly too broad to be foreground lines.  In addition, \OIII/H$\beta$, \Ha\ are all 
ruled out --- as described above, and the small velocity broadening expected for these foreground dwarf galaxies 
(M $\sim$ -17), $\sigma \ls$ 50 km s$^{-1}$, rules out the possibility of broadened \OII.   Three additional \Lya\ 
lines appear to be narrower, although only one appears as narrow as the $\sim$2.5\ang\ instrumental resolution. These 
resemble some of the fainter LAEs from the Subaru-SuprimeCam studies (Kashikawa et al. 2011) and two examples 
from our own Keck-Deimos spectra (MNS3).  For the three found here (labeled 1.9, 2.6, and 3.5 in Figure 2), we rule
out \OII, \OIII/H$\beta$, and \Ha\ (by lack of \OIII\ emission --- see above) as alternative identifications.

In summary, we consider these 13 \Lya\ identifications to be secure, and the foreground 
identifications as well.  

Basic data for each of these 13 LAEs, including position and SNR of detection, are given in Table 1.

\subsection{These are the faintest LAEs yet detected at $z\sim6$}

All 13 LAEs are fainter by a factor of 2 to 5 than the completeness limits of  two Subaru Suprime-Cam 
narrow-band surveys, COSMOS $F \approx 2 \times 10^{-17}$\flux, and the Subaru Deep Field, 
$F \approx 1.6 \times 10^{-17}$\flux\ (see Figure 1 of Takahashi \etal\ 2007).  Although Kashikawa 
\etal\ (2011) include LAEs 2-3 times fainter than these limits, the  detections have large errors, $SNR < 3$, 
with the result that they are drawn from a very incomplete sample.  For this reason,  the faint-end slope of 
the LAE LF is unconstrained by the  Subaru narrow-band data,  as is apparent from the renderings of the LF in 
Kashikawa et al.'s Figures 7 \& 9. Our MNS study has the only sample of LAEs that constrains the faint end 
slope of the LAE LF at $z=5.7$.

The flux level reached in this study is comparable to that achieved by Rauch \etal\ (2008) in their 
study of \Lya-emitters at $2.6 < z < 3.7$, from a heroic 92-hour integration with VLT FORS2.

By accessing the faintest LAEs yet detected at $z\sim6$ we are detecting galaxies that have been previously 
only included in the ionization budget by extrapolation.  Although none of the 13 LAEs discussed here show
a clear continuum flux redward of \Lya, our spectroscopy provides a weaker limit than deep imaging.  MNS3 made 
simulations, based on the $z \sim$ 6 UV continuum LF (Bouwens et al. 2007) and the Stark et al. (2011) equivalent 
width distribution, to estimate that the LAEs in our survey should have $M_{UV} \approx$ -16 to -17 and thus be 
undetected at the depth of HST \emph{Ultra Deep Field}.   Therefore, our LAE sample is an extrapolation of the faintest
LBGs beyond their limit $M_{UV} \sim$ -18.  The \emph{lensed} galaxies from the HST \emph{Frontier Fields} 
are expected to reach LBGs at the depth of the present sample for LAEs.  To detect \Lya\ in that deeper LBG sample 
with the technique used here will probably require the new generation of 30-m telescopes, although deep HST-WFC3 
grism observations might confirm LAEs at this even fainter limit.


\begin{figure*}[t]

\centerline{
\includegraphics[width=7.3in]{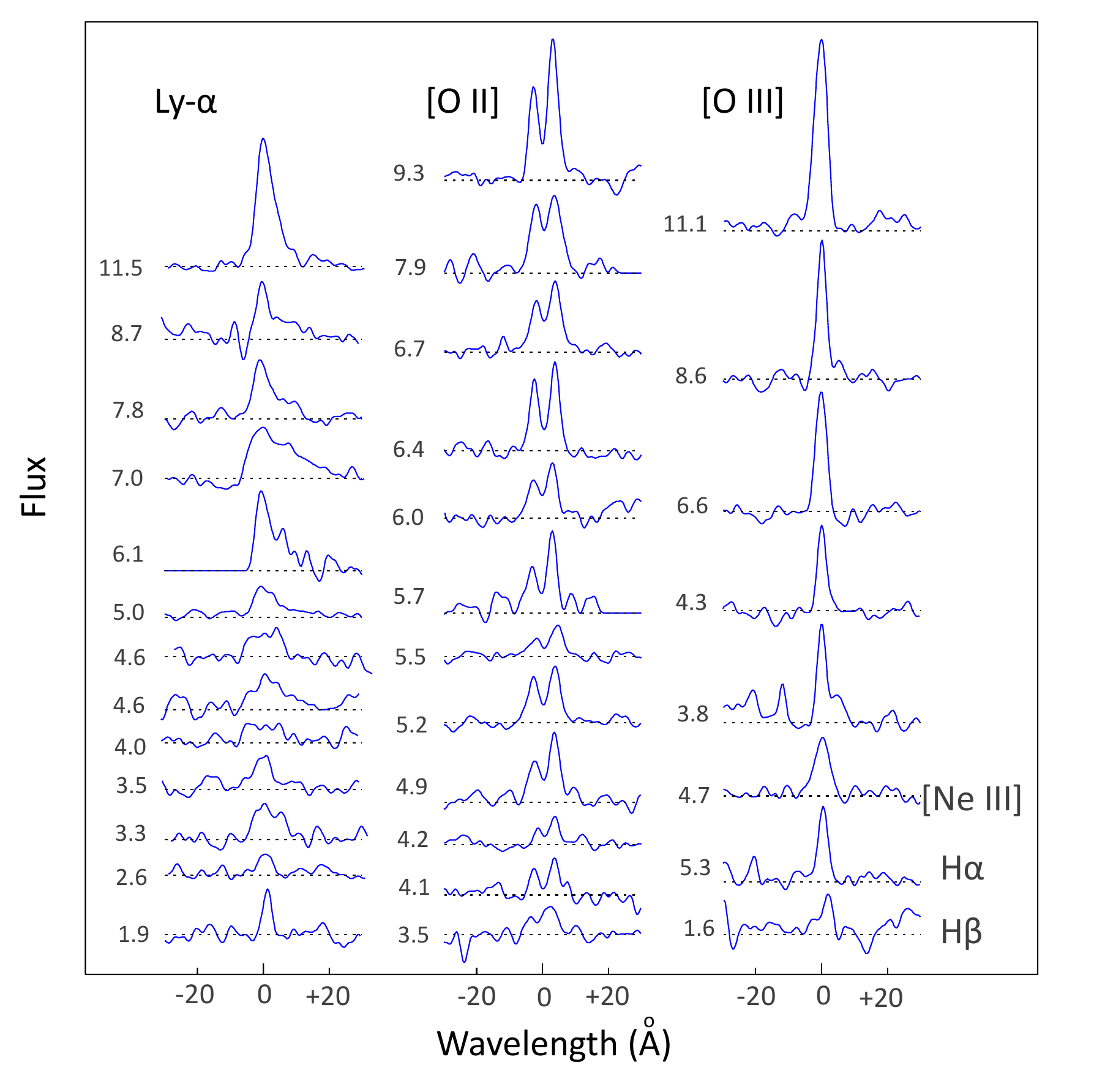}
}

\caption{Spectra of LAE candidates from 2013 and 2014 observations with IMACS
on Magellan-Baade.  The number to the left of each spectrum is its flux in units
of 10$^{-18}$\flux.  The left column shows the 13 detected LAEs at $z\approx5.7$.  The
middle column shows 12 of the 16 detected \OII\ emitters at $z\approx$ 1.20,
which account for $\sim$60\% of the foreground sample.  The four \OII\ detections
not shown are strong signals with fluxes of 6.7, 8.1, 10.7, and 13.2, omitted for clearer
presentation of the remaining spectra.)  The third column shows 8 of the remaining 11 
foreground spectra,  5 of the 8 detected \OIII\ emitters ($z\approx$ 0.64, $\sim$30\% of 
the foreground) and the three remaining  foreground sources, \Ha, \Hb, and [Ne III]. }


\end{figure*}

\section{Measuring the Slope of the LAE luminosity function}

The slope of the LAE LF is the critical determinant of the contribution of low-luminosity 
LAEs to reionizing flux in the early universe.  Uncertainties in the other Schechter function 
parameters --- a lower characteristic luminosity L$^*$ or even a lower space density $\Phi^*$ 
--- are quickly overcome if the slope is steep, $\alpha < -1.5$, needing only a factor-of-two 
more "depth" to reach the photon flux capable of reionizing the universe.  

In  MNS2 we presented a sample of 210 galaxy spectra that showed only a single emission
line in a 140\,\AA-wide search band centered at $\lambda\approx$ 8180\,\AA.   The source 
counts of these candidate LAEs rose rapidly with decreasing flux, but we recognized that most 
of these sources had to be foreground galaxies.  Lacking the high-resolution spectra we now have 
(or any reliably way separate LAEs from foreground), we used published data from Taniguchi 
\etal\ (2007) of foreground [O II], [O III], and H$\alpha$ emitters to remove the foreground statistically, 
leaving a possible LAE LF.  In particular, subtracting our best Schechter-function fits to the 
foreground counts (shown in Figure 8 of MNS2) produced an LAE LF with a faint end slope 
$\alpha\approx$ -2.0 that matched up well with an LAE LF with the same slope from Shimasaku 
\etal\ 2006 (see MNS2, Figure 10) --- one of three acceptable fits to the LAE LF they made using their sample of 
L $\gs$ L$^*$ LAEs.  

Unfortunately, the LAE LF we derived with this method was not unique: our Schechter fits to the LFs of the 
3 foreground populations could not be tightly constrained because they required an extrapolation to 
the faint flux levels of our study: the Taniguchi \etal\ data come from narrow-band imaging observations 
that, like the LAEs, become rapidly incomplete for log F $<$ -17.0. For this reason, we needed to consider 
perturbations on the "best-fitting" foreground LFs to assess the robustness of our result of an LAE LF with 
slope $\alpha \approx$ -2.0.

In that exercise, we learned that we could not rule out a much shallower slope for the LAE LF, even  to 
$\alpha=$ {-1.0}, or even a slope as steep as $\alpha=$ -2.5.  Here we use the term ``realization" to refer
to each of the possible LAE LFs we generate by modifying the foreground LFs within their uncertainties. In 
the process of making such realizations in MNS2, we also found --- not surprisingly --- that the fraction 
of LAEs, LAE/(LAE+foreground) was a sensitive function of the LAE LF slope.  The power of the new data 
presented in this paper is that even a small sample of 40 LAE+foreground sources can greatly reduce 
the range of acceptable realizations.  This is the approach that we now describe.

In MNS2 we adopted the Shimasaku \etal\ Schechter LF fits to their LAE data\footnote{Parameters for 
these LFs are closely matched by the Hu \etal\ 2010 study.} with faint end slopes of -1.0, -1.5, and -2.0, as 
models for the different ``realizations" of the LAE LF we had made by subtracting slightly different levels 
of foreground contamination.  This means that each realization was made to match a Shimasaku \etal\ LF of slope 
$\alpha$, including its L$^*$ and $\Phi^*$ `normalization.'  The Shimasaku \etal\ LFs predict 16-18 bright LAEs 
(log F $>$ -17.0) --- depending on the slope --- over the volume of our survey (see MNS2 Figure 10).  A deficiency 
of that analysis, however, was the graphical, rather than analytical, comparison of our realizations of the LAE LF 
with the Shimasaku \etal\ models (MNS2 Figure 11).  We rectify this here by measuring the steepness of the 
cumulative LAE LF in the Shimasaku \etal\ models, $R$ = $N_{LAE}$(log F > -17.6)/$N_{LAE}$(log F > -17.3) 
--- the ratio of the integrated LAE counts over this flux interval,  and adjusting the foreground LF fits (within 
their uncertainties) to achieve the same quantity for each LAE LF realization.\footnote{$R$ is a proxy for the 
asymptotic slope $\alpha$, which our data --- although well below L$^*$ --- do not reach.}  (Each realization 
also matches the $\Phi^*$ normalization discussed above.)  Table 2 lists the $R$ values for each  model and 
realization and the values of  $\alpha$, log L$^*$, and $\log \Phi^*$ for \OII, \OIII, and \Ha\ foregrounds that were 
used to achieve the match.\footnote{Although cosmic variance of order $\sim$30\% in the $\Phi^*$ 
normalizations of foreground LFs are expected in fields this size, we made changes of only $\Phi^*$ of only 
$\le$10\% from the LFs we adopted from Taniguchi \etal\ (2007).  Beyond this, non-physical LFs result, that 
is, LFs with negative LAEs or diverging with increasing depth.  Cosmic variance is not an issue in this study 
because the foreground LFs had already been measured in one of our fields, COSMOS, and the LCRIS and COSMOS 
fields have similar distributions of number counts versus flux (compare Figures 1 and 2 in MNS2), and also consistent
with the actual measurement of foreground contamination we present here.  We note, however, that the method we 
describe here would work for another field of this size with significantly different levels of the foregrounds, and 
would produce a $z=5.7$ LAE LF with its appropriate cosmic variance of $\sim$20\% for fields of this size.}

If we now calculate  LAE/(LAE+foreground) --- the LAE fraction --- over the interval $F = 2-20\,\10-18$\flux,
for each of these three realizations,  we find values of 0.099, 0.142, and 0.260, corresponding to expected
number of LAEs of approximately 4, 6, and 10, respectively, for a 40-object sample.  These are to be compared 
with the 13 we actually found.  In the next section we describe a simple test of the likelihood of these and
other realizations that, in the end, constrain the allowable realizations to a small range of slopes.

\subsection{The LAE fraction of different realizations and comparison with observations}

In \S3 we discussed the substantial uncertainties in the fluxes of our faint sources.  Even 
at this level of accuracy, the data are probably good enough to fit a Schechter function --- as we did 
in MNS3, but in this paper we use a new method that is robust to photometric errors, measuring 
only the LAE fraction LAE/(LAE+foreground) over a flux interval, $F = 2-20\,\10-18$\flux, and 
comparing this with expectations based on luminosity functions of varying slopes.  This ratio 
is well measured --- despite the uncertainty in fluxes --- because both LAEs and foreground galaxies 
are well bounded, on the faint end by the flux limit of all detections, and on the bright end by L* for 
both LAEs and foreground.   Figure 3 shows that for $F > 2 \times\ 10^{-17}$\flux\ there is only one 
foreground galaxy  out of the total sample of 27, and no LAEs.


\begin{figure}[t]

\centerline{
\includegraphics[width=3.2in]{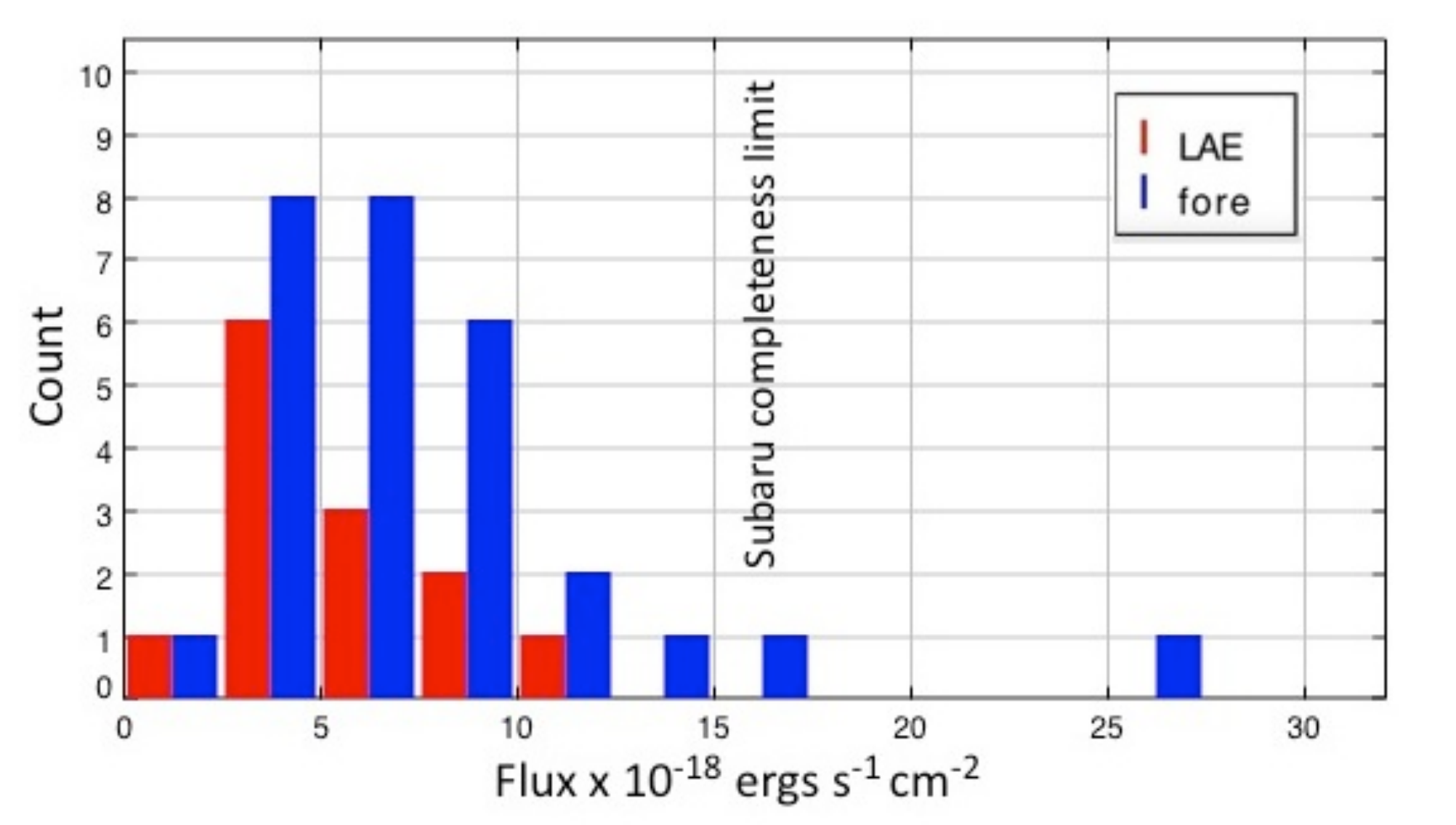}
}

\caption{Flux distribution for LAE and foreground sources, showing the increasing fraction 
of LAE compared to the foreground population with decreasing flux.
\label{fig:LAE_fore_Lum_histogram}}

\end{figure}

Table 3 lists realizations of the MNS2 data we made for the 3 Shimasaku \etal\ models but also spaning the 
full range of plausible faint-end slopes, $-1.5 > \alpha >$ -2.5, in steps of 0.1.   These interpolated Shimasaku \etal\ 
models were generated by quadratic fits to log L$^*$ and log $\Phi^*$ and , each as a function of $\alpha$, based on the 
three models of slope $\alpha =$ -1.0, -1.5, and -2.0.  Table 3 lists these Schechter function parameters for each model 
and its $R$ value, which is compared to the $R$ value of the realization of the data that matches this model. 
For a sense of how much the foregrounds need to be adjusted to produce this full set of models we also include the
$\alpha$, log L$^*$, and log $\Phi^*$ values for the \OII\ and \OIII\ LFs, which account for 90\% of the foreground.  
The progression of these values from $\alpha =$ -1.0 to -2.5 show that, while these values are not ``unique," there is
a predictable manipulation of the parameters that generate the full range of realizations.

With this full range of realizations of  the 2008 data, each matching a Shimasaku \etal\ LAE LF,  we use our new data 
to test the likelihood of each.  This was done by calculating the predicted LAE fraction, LAE/(LAE+foreground), over 
the range $F = 2-20\,\10-18$\flux\, for each realization, and comparing it to the LAE fraction of our new data, 
13/40, or 0.325. 

We use a Monte Carlo test to determine how often the \emph{observed} LAE fraction of 0.325 would be reproduced in 
each of our realizations of the LAE LF --- these results are shown in Table 3.  For example, the LAE LF realization with 
$\alpha_{LAE}$ = -1.0 has an LAE fraction of 0.099 --- 4 LAEs out of 40 total (single-emission-line-only) detections for 
this nearly flat slope.   The Monte Carlo test uses a random draw from a 40 object sample to determine that the observed 
number of 13 LAEs would be found only once in 10,000 trials if only 4 are expected.  This possible LAE LF is therefore ruled 
out.  The slopes -1.5 and -2.0, considered in MNS2, have LAE fractions of 0.142 ($\sim$6 LAEs) and 0.260 ($\sim$10 LAEs), 
corresponding to probabilities of 0.19\% and 22\%, of  finding 13 LAEs.  The ``best fit" of $\sim$40\%  is between 
the LAE LF realizations of $\alpha$ = -2.1 and -2.2.   The likelihood falls for greater slopes: for the realization $\alpha$ = -2.5 
the LAE fraction is 0.477 ($\sim$19 LAEs) and the probability of finding \emph{as few} as 13 LAEs for a 40 object sample has 
decreased to (100-98) = 2\%.  In comparison with MNS2, the flat and modest slopes of -1.0  and -1.5 for the LAE LF --- although 
compatible with the brighter data of Shimasaku \etal\ --- produce too few LAEs and are ruled out by our 13 LAE detections, 
as is the -2.5 slope, which produces too many.  

The result is a probability distribution that is close to Gaussian, with  the mean value of  $\alpha$ = -2.15 and a 
standard deviation of 0.20.  The 2$\sigma$ value is reached at -1.75, as expected, but the  2$\sigma$ on the steep 
side comes in at -2.50 rather than -2.55.  The shot noise associated with this relatively small sample suggests a 
systematic error of $\sim$0.1 in the slope.  

The LAE LF with a faint-end slope of $\alpha$ = -2.15 passes two other tests that show how well it fits the 
data.  The first considers how well the percentage of \emph{each} foreground population in our new data 
compares to values derived in MNS2 for the COSMOS field (where there are data for the foregrounds, as described 
in MNS2), but applied to both of the 2008 search fields.  (In this paper we used that 2008 model as a starting point 
to set the $\Phi^*$ of each foreground.)  With our  best-fitting realization of slope, $\alpha = -2.15$, the 
relative foreground contributions over the log F = -16.8 to -17.7 range are: 57\% for \OII\ compared to 
observed 63\% (1$\sigma$ bounds 42\% - 70\%); 34\% \OIII\ compared to observed 30\% (23\%-45\%); 
and 9\% for \Ha\ compared to observed 7\% (4\% to 16\%).  This validates the foreground model used to 
derive slopes for the LAE LF of $\alpha\sim-2.0$ in MNS2, that is, the parameters for the Schechter functions 
describe the foregrounds well.

The second test concerns the LAE-to-foreground ratio as a function of decreasing flux.
Although we have simply gathered together all the LAEs and foreground galaxies in the flux
interval and focused on a single parameter --- the LAE fraction, we can learn something from Figure 3 about the 
distribution --- the increasing fraction of LAE/foreground with decreasing flux.  Again, the 
best fit LAE LF derived with the new data is in agreement this observed trend: brighter than 
log F = -17.0, foreground galaxies in the model outnumber LAEs by 9 to 1.   At log F = -17.0 
this ratio has dropped to 4.4 to 1, and at log F = -17.6\flux\ LAEs are almost one-to-one with
the foreground.  For all the uncertainty in the fluxes, this is what the data of Figure 3 show.

\subsection{Comparison with the Keck-DEIMOS results}

In MNS3 the results of Keck-Deimos observations in 2011 and 2012 were presented and analyzed, 
including the first recovery of faint LAEs in MNS, 6 LAEs with fluxes between $F = 5 - 10~ \10-18$\flux.  
A maximum likelihood technique was used to find a LF faint-end slope of $\alpha\sim-1.7$, shallower 
than found here, but the $\alpha\approx$ -2.0 slope found in MNS2 using a statistical correction of the 
foreground contamination is within the 1$\sigma$ uncertainty of the both MNS3 and the present result.  
The methodology used here to measure the faint-end slope is not easily applied to the MNS3 data, since
there was a prioritization of DEIMOS targets --- based on previous low-resolution IMACS spectroscopy ---
that favored objects that were narrowed-down to be either LAE or \OII\ foreground over those without 
additional information following the original detection in the 2008 search data (see MNS3).   Also, the 
LAEs found in MNS3 cover only the brighter part of this paper's sample (see Figure 3), which means that 
the LAE fraction is expected to be smaller, 26\% instead of 33\% --- according to the best-fit model we 
find here.  Still, it  appears that the result of MNS3 points to a flatter slope.   We stress, however, that the 
derivation of a probable $\alpha$ = -2.15 slope in this study is completely compatible with the data and 
analysis of MNS3.

A strength of the present work is that the LAE and foreground spectra represent a nearly complete 
($\sim$85\% of targeted objects) sample, randomly selected by the spatial constraints of the multislit 
mask technique, that should be unbiased.  This simplifies the analysis here.  The unbiased selection of 
targets, and the much larger sample of confirmed LAEs, makes the present work the best assessment of 
the faint LAE population to-date, providing the strongest constraint on the faint-end slope $\alpha$ of 
the LAE LF.

\section{The LAE population provides a substantial fraction of reionizing photons}

At $z=5.7$, our sample lies past the redshift of full reionization at $z=6.0$ --- our LAEs 
contribute to \emph{maintaining} ionization by balancing recombination.  However, because
$z=5.7$ and $z=6.0$ are separated by only  by 64 million years, and by an additional 200 Myr 
to $z=7$, it is reasonable to  believe that our sample is representative of  the similar emission-line
galaxies \emph{within} the reionization epoch.  Furthermore,  since HI absorption seems to 
substantially attenuate the \Lya\ signal at $z\gs7$ (see \S1), observing LAEs at $z=5.7$ may turn 
out to be the best epoch to study the properties of LAEs at earlier times. 

In MNS2 we reviewed a number of issues that were related to the possibility that the
faint-end slope of the LAE LF is steep, $\alpha\approx$ -2.0, something that the present
study confirms.   The identification of these faint LAEs as systems of halo mass 
$10^{10} - 10^{11}\, \Msun$, at a space density equivalent to several objects per today's 
L$^*$ galaxy, motivated our contention that these are the likely progenitor components 
of L$^*$ galaxies, and that these lower mass systems are the probable source of the metal 
enrichment of the IGM at this early epoch.   In this connection, the resolved profiles of most 
of the \Lya\ sources presented in this paper have rest frame widths of several hundreds of 
kilometers per second, a necessary though not sufficient condition for ascribing a large 
outflow velocity, which additionally requires an as-yet-unmeasured local-standard-of-rest.

Here we consider only the ramifications of this now well-measured faint-end slope of the LAE LF at 
$z=5.7$ for the question of the sources of reionization of the IGM.  In a recent study motivated in part 
by the results presented here, Gronke \etal\ (2015) predict a Schechter-like slope of the LAE LF based 
on a model that uses the UV-LF of LBGs and the distribution of their \Lya\ strengths as a function of 
UV-luminosity and redshift (see also Garel \etal\ 2015).  Gronke \etal\ predict a faint-end slope of 
$\alpha <$ -2.0 for the LAE LF at z $>$ 4 --- somewhat steeper than the LF for LBGs galaxies --- and 
that this slope holds until a turnover around $10^{40}$\lum $<  L <  10^{41}$\lum.  These predictions 
of a steep slope of the LAE LF in agreement with the measurement reported here, and its continuation to 
$L < 10^{41}$\lum, bodes well for our argument that LAEs play a significant, perhaps even a dominant 
role in the  reionization of the early universe, as we now show.

In MNS1 and MNS2  we derived the star formation rate (SFR) density required to maintain ionization at 
$z\sim5.7$ from the LAE LF flux.  The uncertain parameters for calculating this quantity are the production 
rate and escape fraction of \Lya\ and LyC photons and the clumping factor  of the IGM.   MNS1 derived the 
equation for the critical luminosity density $\mathcal{L}$ in \Lya\ required to maintain ionization at $z=5.7$,

\begin{equation*}
\begin{split}
(1) \,\, \mathcal{L} = 3.0 \times 10^{-40}\, ergs^{-1}\, s^{-1}\, Mpc^{-3} \times \zeta  \times \left(\dfrac{1+z}{6.7}\right)^3 \left(\dfrac{\Omega_b h^2_{70}}{0.047}\right)^2
\end{split}
\end{equation*} 

where 
\begin{equation*}
(2)\,\,\,\,\,\,\,\, \zeta = C_6 (1 - 0.1f_{Lyc,0.1}) \left(\dfrac{f_{Ly\alpha,0.5}}{f_{Lyc,0.1}}\right)
\end{equation*}

\noindent combines the clumping factor, the \Lya\ escape fraction, and the LyC escape fraction, 
normalized to values of 6, 0.5, and 0.1, respectively.  A value of $\zeta\approx 1$ represents current estimates of
these values.


\begin{figure}[t]

\centerline{
\includegraphics[width=3.2in]{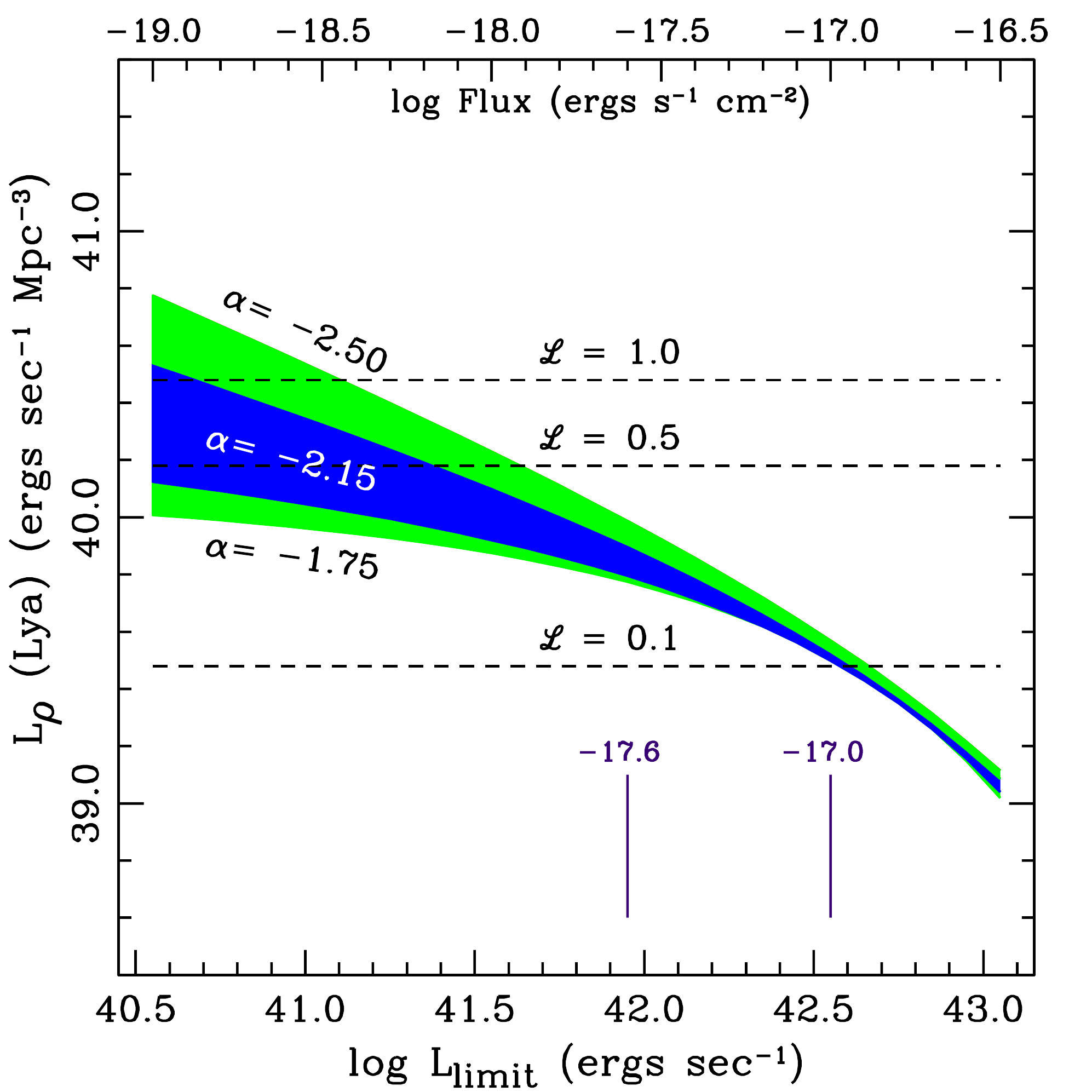}
}

\caption{Level of luminosity density required for maintaining reionization at $z\approx5.7$ with a population 
of faint LAEs.  The blue and green shadings show the $\pm1\sigma$ and $\pm2\sigma$ bounds of a best-fit 
slope is $\alpha = -2.15$ and $\sigma = 0.20$.  Within these limits, there is substantial progress toward reaching 
critical flux density, $\zeta=1$: 22\% is reached at the flux limit of our observations, 35\% if the LF continues
to a factor-of-three-fainter flux limit, and $\sim$56\%  if it continues a full factor-of-ten.  If the LyC 
escape fraction reaches as high as 20\%, the full reionizing budget could be reached at that point.  Such a higher 
escape fraction is consistent with trends of increasing redshift and decreasing luminosity found in lower-redshift 
samples, increasing the likelihood that LAEs alone can provide the critical flux density to complete reionization at 
$z\sim6$.
\label{fig:reionization_flux_density}}

\end{figure}

The rising LF for faint LAEs we have confirmed here is an important step towards showing that galaxies at $z=5.7$ 
are capable of maintaining ionization and, by implication, that a similar population of low-mass, low dust galaxies 
made a substantial contribution to reionization at $z\gs7$, only $\sim$300 Myr earlier.  As discussed in \S3.2, 
the faintest LAEs in our sample are likely to have $M_{UV} \approx$ -16 to -17, thus they add to the fraction of
reionizing flux found for the LBG population, whose limit is presently $M_{UV} \sim$ -18.

In Figure 4 we reframe Figure 12 of MNS2 with the new limits on the faint-end slope, confirming that such systems
played a substantial role in the ionization of the IGM.  Figure 4 shows the luminosity-density in \Lya\ as a function of the 
limiting luminosity of the LAE LF that has been measured.   The critical flux density, $\mathcal{L}$, from Eqn. (1), is shown 
for 10\%, 50\%, and 100\% of the flux required for full ionization.  The blue shaded region shows how 1$\sigma$ limits on 
the faint-end slope map onto the reionization flux.  Assuming a modest factor-of-three extrapolation in limiting luminosity 
of our faint-end slope, our observations already reach a level of $\sim$35\% of the critical density, and a factor of three 
further extrapolation brings us to the $\sim$55\% level.

A faint end slope $\alpha > -2.0$ is unphysical, of course, in the sense that extrapolation of this LF indefinitely
is unbounded.  However, there is no reason to suspect that the physics responsible for the steep interval found
here --- significantly steeper than for any lower-redshift sample of galaxies, continues to apply.   A shallowing
of the slope $\alpha$ or even a cutoff for much fainter LAEs would not be unexpected.  Furthermore, the steep slope 
we find for the LAE LF may not be entirely due to a steep increase in the actual number of objects, since an increasing 
\Lya\ escape-fraction, also not unexpected in lower-luminosity (lower-mass) systems (Schaerer \etal\ 2011), could 
be partly responsible.  Finally, we note that Figure 4 uses an escape fraction of LyC photons of only 
10\%, a conservative value that may also increase with higher redshift and lower-mass systems (Hayes \etal\ 2011; 
Blanc \etal\ 2011; Nestor \etal\ 2011, 2013; Dijkstra \& Jeeson-Daniel 2013; Jones \etal\ 2012; Jones \etal\ 2013; 
Cassata \etal\ 2014).  If so, reaching the full flux needed to maintain or drive reionization may be achieved with 
a continuation of the steep LF for only a factor-of-ten beyond the luminosity range covered in this study.  

Finally, we note a recent study by Topping and Shull (2015) that suggests a boost in production efficiency --- 
LyC photons per unit SFR --- based on new models of rotating hot stars (see also Leitherer \etal\ 2014).  
Although we have used previous estimates of LyC production efficiency to facilitate comparisons with previous 
work, such changes would push the contribution of LAEs to reionization that much closer to, or above, 
the critical SFR density.  

\section{Conclusion}

We have confirmed a steep faint-end slope of the luminosity function of Lyman-$\alpha$ emitters  at $z = 5.7$ 
by finding a $\sim$32\% fraction of LAEs in a sample of 42 extremely faint emission-line galaxies.  A robust test 
shows that this fraction of LAEs is inconsistent with faint-end slopes much flatter than $\alpha$ = -1.90, and that 
a slope of $\alpha$ = -2.0 or greater has a high probability.  A slope this steep suggests a substantial, perhaps 
dominant contribution by LAEs to maintaining reionization at this epoch, with a moderate extension of the $\alpha 
\approx$ -2.0 slope by a factor of $\sim$10 or to fainter systems needed account for much or even all the required 
flux.  Considering the proximity in time of these LAEs to objects within the reionization epoch, it is reasonable to 
imagine that similar emission-line galaxies at $z >7$ make a substantial contribution to reionization in the early universe.

\section{Ackowledgments}

AH is supported by an appointment to the NASA Postdoctoral
Program at the Goddard Space Flight Center, administered by Oak Ridge Associated Universities through a contract with NASA.


\clearpage
\vspace{0.1in}
\appendix

The line detections shown in Figure 2 have been smoothed with a gaussian kernel of width
$\sigma =$ 1.0\ang\ and are shown centered on the line in an interval of 60\ang.  These
choices make it more difficult to judge of the reality of the line with respect to sky noise and its 
prominence compared to other possible features.  To remedy this, we replot in this Appendix the 
same spectra of Figure 2 for 13 LAEs and 12 \OII-emitters over the full $\sim$135\ang\ bandpass 
of the search window, 8115\ang\ to 8150\ang.  The smoothing has been reduced to $\sigma =$
0.387\ang\ (1 pixel, compared to the instrumental resolution of $\sim$2.5\ang\ FWHM) to show the noise 
after sky subtraction.  The vertical scale for the spectra plotted in black is  marked at the bottom left 
at the level 33 counts per pixel, equivalent to a flux of $1.5 \times 10^{-19}$ \flux.  The stronger 
spectra plotted in blue are shown at half this scale and marked accordingly.   

Figure 5 confirms that these emission-line sources are the same objects found in the  2008 search, 
by demonstrating that each is the strongest feature in the band of the search.  That is, in addition 
to  detecting an emission line within 10\ang\ of search detection --- as shown in Figure 1, that line 
is also the strongest  feature in the bandpass.  Identification of these emission lines with those of 
the search are based on a coincidence of sky  position to $\sim$0.5 arcsec --- the placement of the 
slit for the confirmation spectrum, the spatial position along the slit to within $\pm2$ arcsec, and 
to a correspondence of the strongest feature in the confirmation spectra to the predicted wavelength, 
with a typical agreement of 5\ang.

Moreover, the identified lines are the only {\it statistically significant} features in the spectra.  We have 
calculated the signal-to-noise ratio (SNR) of each emission feature using the unsmoothed, 
sky-subtracted spectra at their raw dispersion of 0.387\ang\ pixel$^{-1}$.  For each object we
selected the pixels over which the line flux had been measured, marked in Figure 5 by the short
red lines.  The remaining pixels were used to determine the noise in counts, without removing
possible additional sources.  (Segments that are set to zero are gaps in the CCD mosaic array.) The 
SNR for each feature, determined as the the square root of the sum of the squares of signal-to-noise 
calculated pixel by pixel,  is recorded next to each feature: each line is a highly significant detection 
with $SNR > 6$. These determinations of SNR are in good agreement with those done for the 2008 search 
data (see Figure 3 of MNS2), which were determined photometrically using the two-dimensional spectral 
images (``spaxels'').  It is clear from inspection of the marked lines and the SNR that no other features 
are detected over these wavelength intervals at a significance of over 5$\sigma$, the standard criterion 
for a detection in photometric or spectroscopic data.

Returning to the question of source identification discussed in \S3.1, we now review the possibility
of misidentification of what are clearly real sources.  Spectra with multiple, well-spaced lines are clearly 
not LAEs, so these cases in the right column of Figure 2 have not been replotted Figure 5.  Only [O II] can 
be confused with \Lya. At the full 2.5\ang\ FWHM resolution of these spectra, it is clear that the \OII\ doublet
($\lambda\lambda$3726, 3729\ang) can be easily distinguished from \Lya, even in cases of relatively low SNR.  
Each of the faintest 6 sources identified here as \Lya\ have sufficient SNR to distinguish them from \OII\ 
because they do not show any structure with the 5\ang\ (redshifted) spacing of the \OII\ doublet.  We 
consider the most ambiguous  case to be the faintest \OII\ emitter, which could be faint \Lya\ with a noise 
spike on the to the blue that has the proper 5 \ang\ spacing.   Judging from the  SNR, it appears that this is 
a possible but unlikely (probability < 10\%), similar to the chance that any one of the five faintest \Lya\ 
lines is actually a very noisy \OII.

In summary, inspection and analysis of the spectra of \Lya\ candidates at full spectral resolution confirms 
that, with high probability, the lines recovered in the 2013 \& 2014 observations are those found in the
2008 search, and that the discrimination between \lya\ emission and foreground sources is secure.

\begin{figure}[t]

\centerline{
\includegraphics[width=6.3in]{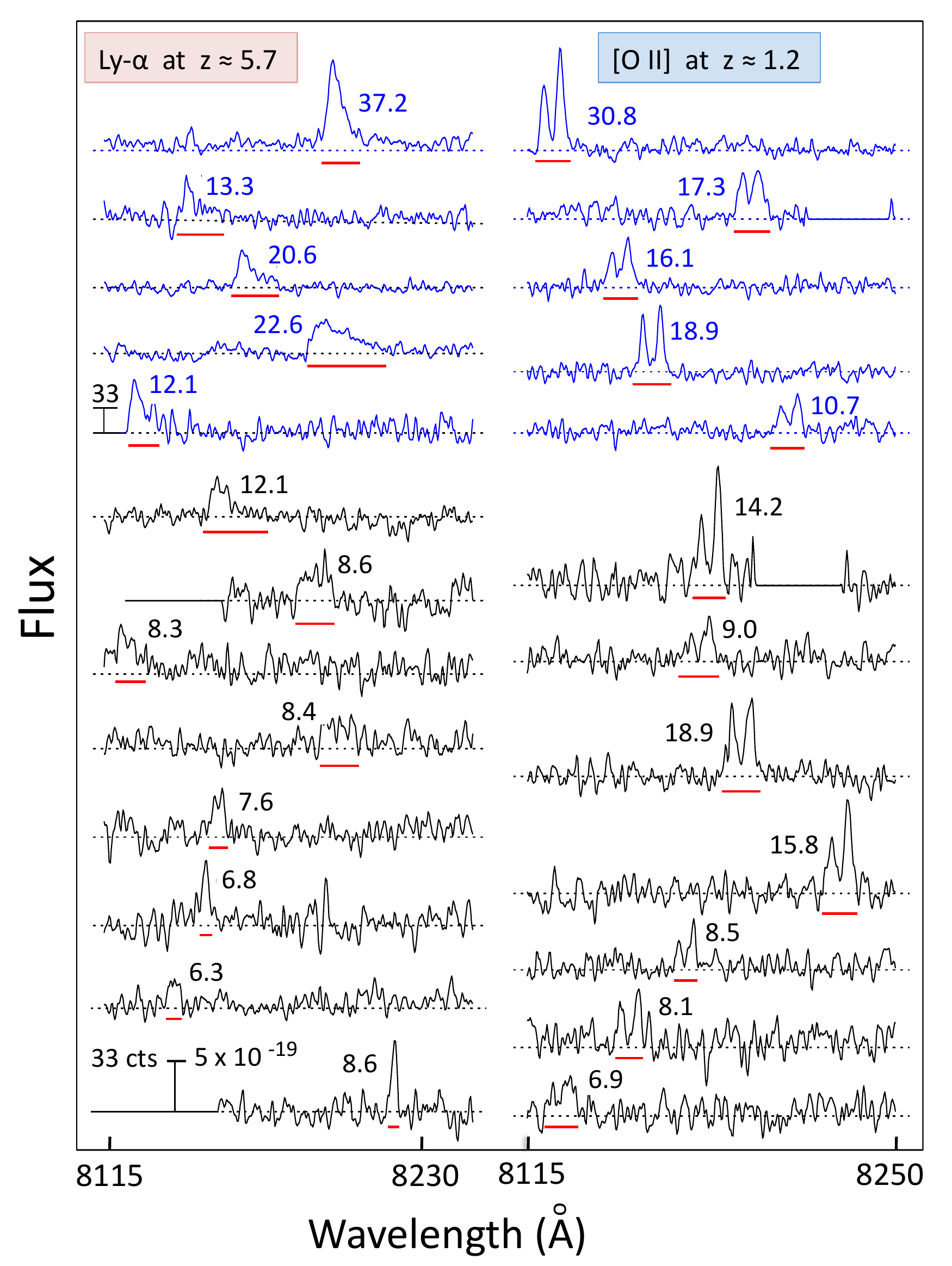}
}

\caption{Sky-subtracted spectra of identified \Lya\ lines (left) and foreground \OII-emitters (right), smoothed with 
a gaussian kernel of width $\sigma =$ 1 pixel (0.387\ang), and plotted over the full interval of the narrow-band filter, 
8115--8250\ang . (These are the same objects shown in Figure 2 left and middle columns.)   The vertical scale, in counts 
and equivalent flux (ergs s$^{-1}$ cm$^{-2}$), is shown at the bottom left for spectra in black; the scale of spectra plotted 
in blue is compressed by a factor of two, as shown (5th from top, left-side).  The zero-level of each spectrum is shown 
as a dashed line; solid flat lines are for chip gaps in the CCD mosaic array.  Detected \Lya\ and \OII\ are marked by 
short red lines.  The numbers adjacent to each feature record the SNR of the feature, computed from pixel-by-pixel 
signal-to-noise, as described above.  All features well exceed to customary 5$\sigma$ criterion for detection.}

\label{fig:LAE_OII_spectra}

\end{figure}

\clearpage

\begin{table*}
\begin{center}
{\scriptsize
\caption{Identified LAEs at $z=5.7$}\label{LAEs}}

\begin{tabular}{lcccccc}

\tableline\tableline

\\  \#  &   Identification &   RA  &  DEC      &   $\lambda_{\Lya}$   &  Flux $\times 10^{18}$     &   SNR
\\  \,\,\,\,  &      &        (2000.0)     &   (2000.0)     &   angstroms  &  ergs\,\,s$^{-1}$\,cm$^{-2}$         
\\      
\tableline

\\  \, 1    &   14.5+3-0.91   &   15:23:00.333 & -00:13:28.18   &   8199  &   11.5    &   37.2
\\  \, 2    &   17.5-2-0.18    &  15:23:12.900  & -00:15:33.56   &    8196  &     8.7    &  13.3
\\  \, 3    &   54.5+6-0.43   &  15:23:08.617  & -00:01:24.37   &   8146   &     7.8    &  20.6  
\\  \, 4   &    58.5+5-0.94   &  15:23:18.792  & -00:02:31.55   &   8167   &     7.0    &  22.6
\\  \, 5   &    40.5-4-0.95    &  15:23:43.210  & -00:15:01.20   &   8127   &     6.1    &  12.1
\\  \, 6   &     31.5+6-0.66   &  15:22:56.159  & -00:06:23.60   &   8157   &     5.0   &  12.1
\\  \, 7   &     44.5+4-0.54   &  15:23:12.015  & -00:05:46.11   &   8192   &     4.6   &    8.6
\\  \, 8   &     63.5+8-0.92   &  15:23:08.087  & +00:01:53.87   &  8121   &     4.6   &    8.3
\\  \, 9   &     13.5+3-0.45   &  15:22:58.690  & -00:13:24.69    &  8201   &     4.0   &    8.4
\\ 10   &   55.5-5-0.43   &  15:23:57.310  & -00:13:16.30    &  8157   &     3.5       &   7.6
\\ 11   &   32.5-2-0.74    &  15:23:27.052  & -00:13:47.32    &  8147   &     3.3      &   6.8
\\  12  &  46.5-5-0.75    &   15:23:51.911   & -00:15:05.34   &  8141   &     2.6      &   7.3
\\  13  &  56.5-4-0.68    &  15:23:53.999   & -00:12:04.48   &  8221   &     1.9       &   8.6
\\
\\
\tableline
\end{tabular}

\end{center}
\end{table*}

\begin{table*}
\begin{center}
{\scriptsize
\caption{Shimasaku \etal\ LAE LF models and MNS2 LAE LF realizations }\label{MNS2 models and realizations}}

\begin{tabular}{lcccccc}

\tableline\tableline

\\     model LF$_{LAE}$   &   $R_{model}$  &  $R_{realization}$    &   &  Foreground LF parameters    &    &

\\  \,\,\,\,  $\alpha, L^*, \Phi^*$    &  N(-17.6)/N(-17.3)     &  N(-17.6)/N(-17.3)    &     \OII  & \OIII  &     \Ha  
\\      
\tableline

\\ \, -1.0,   42.72,    -2.92     &     1.621     &     1.628  &  [-1.44,  41.47, 0.768] & [-1.69, 41.42, -0.172] & [-1.69, 41.49, -1.392]    
\\ \, -1.5,   42.90,    -3.20     &     1.895     &     1.891  &  [-1.39,  41.52, 0.790] & [-1.68, 41.42, -0.063] & [-1.67, 41.49, -1.287]
\\ \, -2.0,   43.20,    -3.80     &     2.284     &     2.286  &  [-1.30,  41.48, 0.836] & [-1.60, 41.42, -0.017] & [-1.60, 41.49, -1.303]
\\
\tableline
\end{tabular}

\tablecomments{(1) LF$_{LAE}$ Schechter function parameters from Shimasaku \etal\ (2006); (2) ratio of integrated LAE counts, N(log F $>$ -17.6\flux)/N(log F $>$-17.3\flux), for Shimasaku \etal\ model, and (3) for MNS2 data realization; 
(4) Foreground LFs: Schechter parameters [$\alpha$, log L$^*$, log $\Phi^*$]}
\end{center}
\end{table*}

\begin{table*}
\begin{center}
{\scriptsize
\caption{LF function fits and probabilities}\label{LF fits}}

\begin{tabular}{lccccccccc}

\tableline\tableline

\\     model LF$_{LAE}$   &  $R_{model}$  &  $R_{realization}$    &  [O II] LF &  [O III] LF & LAE fraction   & LAEs  &  Monte Carlo     &  Probability   &    
\\  \,\,\,\,  $\alpha, L^*, \Phi^*$    &     &    & $\alpha, L^*, \Phi^*$    & $\alpha, L^*, \Phi^*$    &    \,\,\,\,\  &   &   $n = 13$    & $\%$  &        
\\      
\tableline

\\ -1.0,   42.72,    -2.92     &     1.621     &     1.628  &  -1.44, 41.47, 0.768      &   -1.69, 41.42, -0.172   &    0.099     &     4.0   &   1.0E-4   &   0.01\%
\\ -1.5,   42.90,    -3.20     &     1.895     &     1.891  &  -1.39, 41.52, 0.790      &   -1.68, 41.42, -0.063   &    0.142     &     5.7   &   1.9E-3   &   0.19\%
\\ -1.6,   42.95,    -3.30     &     1.959     &     1.957  &  -1.38, 41.48, 0.796      &   -1.65, 41.42, -0.057   &    0.151     &     6.0   &   3.2E-3   &   0.32\%
\\ -1.7,   43.01,    -3.40     &     2.029     &     2.027  &  -1.37, 41.48, 0.804      &   -1.64, 41.42, -0.049   &   0.177     &     7.1   &   0.015     &   1.5\%
\\ -1.8,   43.07,    -3.52     &     2.105     &     2.107  &  -1.35, 41.48, 0.812      &   -1.62, 41.42, -0.041   &   0.198     &     7.9   &   0.039     &   3.9\%
\\ -1.9,   43.13,    -3.65     &     2.171     &     2.171  &  -1.34, 41.48, 0.822      &   -1.64, 41.42, -0.031   &   0.218     &     8.7   &   0.076     &   7.6\%
\\ -2.0,   43.20,    -3.80     &     2.284     &     2.286  &  -1.30, 41.48, 0.836      &   -1.60, 41.42, -0.017   &   0.260     &    10.4  &   0.224     &   22\%         
\\ -2.1,   43.28,    -3.95     &     2.392     &     2.391  &  -1.21, 41.41, 0.754      &   -1.58, 41.36,  0.063   &    0.285     &    11.4  &   0.345     &   35\%
\\ -2.2,   43.36,    -4.12     &    2.506      &    2.508   &  -1.17, 41.41, 0.756      &   -1.55, 41.36,  0.081   &    0.302     &    12.1  &   0.550     &   45\%
\\ -2.3,   43.45,    -4.30     &    2.637      &    2.639   &  -1.16, 41.41, 0.712      &   -1.47, 41.36,  0.115   &   0.363     &    14.5  &   0.741     &   26\%
\\ -2.4,   43.54,    -4.50     &    2.783      &    2.781   &  -1.11, 41.41, 0.680      &   -1.40, 41.36,  0.147   &   0.416     &    16.6  &   0.907     &    9.1\%
\\ -2.5,   43.63,    -4.71     &    2.944      &    2.944   &  -1.05, 41.38, 0.593      &   -1.18, 41.34,  0.236   &   0.477     &    19.1  &   0.981     &    2.0\%
\\
\\
\tableline
\end{tabular}

\tablecomments{(1) LF$_{LAE}$ Schechter function parameters based on Shimasaku\etal\ (2006); (2) ratio of integrated
LAE counts, N(log F $>$ -17.6\flux)/N(log F $>$-17.3\flux) for Shimasaku \etal\ model, and (3) for MNS2 data realization;  
(4) LF [O II] Schechter function parameters; (5) LF [O III] Schechter function parameters; (6) predicted LAE fraction, 
LAE/(LAE+fore), over flux interval -17.6 $>$ log F $>$ -17.3; (7) expected number of LAEs; (8) fraction of cases in Monte 
Carlo test with 13 LAEs and 27 foreground galaxies; (9) Probabiliy of LF$_{LAE}$ slope $\alpha$. }
\end{center}
\end{table*}

\end{document}